# Oscillations in two-person avoidance control

Lazar L. Kish

*Department of Physics, University of Illinois at Urbana–Champaign, 1110 West Green Street, Urbana, IL 61801-3080, USA*

**Abstract**

Social interaction dynamics are a special type of group interactions that play a large part in our everyday lives. They dictate how and with whom a certain individual will interact. One of such interactions can be termed "avoidance control". This everyday situation occurs when two fast-walking persons suddenly realize that they are on a frontal collision course and begin maneuvering to avoid collision. If the two walkers' initial maneuverings are in the same direction that can lead to oscillations that lengthen time required to reach a stable avoidance trajectory. We introduce a dynamical model with a feedback loop to understand the origin and properties of this oscillation. For the emergence of the oscillatory behavior, two conditions must be satisfied: i) the persons must initiate the avoidance maneuver in the same direction; ii) the time delays in the feedback loop must reverse the phase of the players' positions at the oscillation frequency. The oscillation can be terminated at any time if one of the walkers decides to stop and/or to communicate. By taking over the control of the situation the walker cuts the feedback loop. Similar oscillatory situations may potentially cause major collisions between autonomous vehicles and airplanes with airborne communications, but in autopilot mode.

## 1. Introduction

The modeling of conflict in dynamical control systems subject to multiple controlling agents, or the field of differential games, has been intensively examined in literature since its origin with Rufus Isaacs in 1965 [1]. A classic example is the homicidal chauffeur problem, which features a pursuing and an evading agent. Leitmann and Skowronski [2] consider a generalization of such systems where one of the agents seeks to evade the other, while the second one moves with a prescribed set of possible trajectories; here, no two-way interaction (feedback loop) exists. The analysis of such problems is potentially applicable to any area featuring collision prevention as an objective, such as maritime and avionic avoidance control [3]. In this paper, we examine a dynamical system in which both agents, initially on a collision course, seek avoidance of the other. We show that, within certain parameters, feedback inherent in this system is supportive of oscillation. The presence of this oscillation increases the time necessary for the two agents to escape collision, and thus increases the time requirement of the avoidance maneuver. Similar oscillatory situations may potentially cause major collisions between autonomous vehicles or cars in autopilot mode, especially given that avoidance problems have already lead to two fatal accidents with the Tesla car in autopilot mode [4,5]. Another obviously relevant topic is the emerging aviation technology *Airborne Communications* [6,7] where malfunctioning avoidance control can potentially lead to major accidents.

## 2. Model

First, to illustrate the mechanism leading to oscillations in avoidance control we use a simplified two-dimensional model. Consider the situation where two persons, Alice and Bob, are walking towards each other on a collision course with paths parallel to the *x* axis (see Figure 1) at coordinates $\vec{r}_A(t)$ and $\vec{r}_B(t)$. We assume that Alice and Bob realize the danger of collision at the same time. When this happens, in order to avoid collision, they will initiate avoidance maneuvers by adding correcting velocity components ($v_A^y$ and $v_B^y$) in the *y* direction. The conditions for avoidance are determined by multiple factors, including the widths of the persons, their distance and velocity.

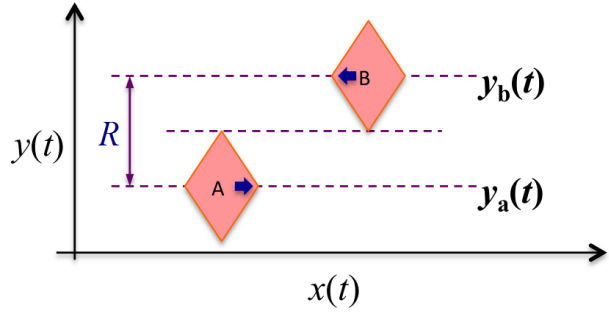

Figure 1. The generic situation. Alice (A) and Bob (B) are proceeding along the y axis in the opposite direction. When the difference of the $y$ coordinate of their center is less than $R$, they are on a collision course and an avoidance control maneuver is required.

For the purposes of our model, we combine these factors into a single parameter $R$, called the *collision radius*. Given that the two objects move parallel to the x axis, Alice and Bob are on a collision course when the following inequality is satisfied:

$$|y_A(t) - y_B(t)| < R \ . \tag{1}$$

When the $y$ components of the initial coordinates of Alice and Bob are identical, $A_y = B_y$, then they are on a *frontal collision course* and the signs of $v_A^y$ and $v_B^y$ are chosen randomly, with an assumed 0.5 probability. First, suppose, the correcting velocity components satisfy $|v^y|_A = |v^y|_B = constant$. If Alice and Bob choose the same sign for their correcting velocity components, this first step of the avoidance maneuver is unsuccessful because they are on a new frontal collision course. If during the next maneuvers their random choices remain identical they will execute synchronized oscillations, as can be observed during certain everyday situations.

In general, the collision avoidance is a highly complex phenomenon with a number of dynamical parameters in R and the control of all velocity components during the maneuvering. Factors include differences between Alice and Bob in their judgments, time-delayed decisions, actions, and in the components $v_A^y$ and $v_B^y$, as well as modulation of the x-components of the velocities. Moreover, the perceived collision radius $R(...)$ can not only be asymmetric between Alice and Bob but also depends on a number of variables such as speed, distance, boundary conditions, cognition characteristics and even the personality of the walker.

In this paper, our goal is to set up and analyze a "minimalist" model of human avoidance control that captures the essence of the mechanism of oscillations. We will see that confinement in the $y$ direction can enhance the lifetime of oscillations. Conversely, asymmetries in the properties of Alice and Bob, such as differences in physiological and psychological responses (e.g. time delay, correcting velocity, decision statistics) can assist the termination of this synchronicity.

To simplify our model, we have quantized the inhabitable y-positions to a discrete set. Two scenarios were constructed within this model. In the first case, that is a binary model we have restricted the inhabitable y-positions to two possible values. The second scenario represents an increase in complexity with a ternary model including three possible values. First we address the time delays in these discrete models.

*2.1 Time delays in the feedback loop*

For the following discussion we will assume that the time-delay parameters of Alice and Bob can be combined into single parameters $\alpha_A$ and $\alpha_B$. This leads to an effective feedback loop in the positions where the position of Alice $\vec{r_A}(t)$ and the position of Bob $\vec{r_B}(t)$ is converted by some function into the position of Bob, after a time delay of $\alpha_B$, and vice-versa for Alice.

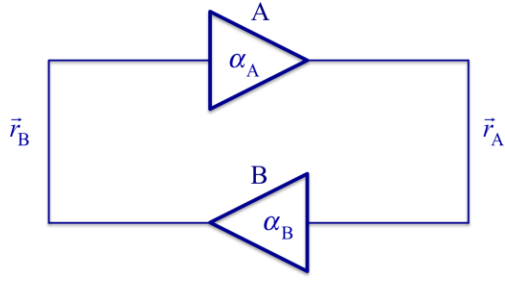

Figure 2. The simplified feedback loop and the times delay in the discrete models.

*2.2 Binary model*

Within our first scenario, the agents are constrained to move in the opposite directions on one of two antiparallel paths: along $y = y_+ = h$ or $y = y_- = -h$. Physically, this mimics a situation where space is limited, as in the case of two persons meeting in a narrow corridor which is just wide enough for two. The geometry of this scenario is illustrated in Fig. 3.

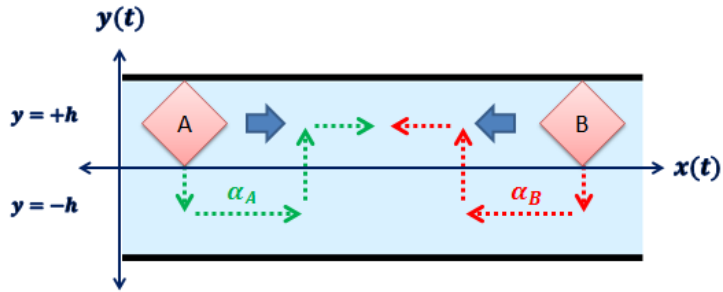

Figure 3. An illustration of the discrete binary avoidance control model with two possible paths of motion, mimicking the real life situation of two persons in a narrow hallway attempting bypass each other.

If the agents possess at all times a perfect awareness of the situation (without lag), and can act instantaneously in response, it is reasonable in the "narrow hallway" limit of this model to suppose that suitable avoidance trajectories will be determined by their relative distance in the $y$-direction:

$$\Delta y = |y_A - y_B| \qquad (2)$$

When this quantity is smaller than the collision radius, $\Delta y < R$, each agent will perform an avoidance control maneuver by changing their position to the other available path. In this limit, they would enter an infinitely fast oscillation.

However, in the real world thoughts and actions cannot occur instantaneously. To make our model more realistic, we introduce time delays into the actions of Alice and Bob. One way in which we can do this is by restricting subsequent actions by Alice or Bob to occur only after respective time intervals of $\alpha_A$ and $\alpha_B$. Then we can make the avoidance control parameter a time average of the relative distances over the $\alpha$ time delay rather than its instantaneous value. This gives us the integral

$$\frac{1}{\alpha}\int_{t-\alpha}^{\alpha} |y_A[t] - y_B[t]| \qquad (3)$$

After every time interval of $\alpha_A$ or $\alpha_B$, Alice and Bob compare the value of this quantity to the collision radius $R$.

If the evaluated average difference is smaller than the collision radius as below, then the agent will perform an avoidance action.

$$R > \frac{1}{\alpha}\int_{t-\alpha}^{\alpha}|y_A[t] - y_B[t]| \tag{4}$$

Let us set the agents on a collision course by initially requiring them both to occupy $y(t=0) = y_+$. If both agents possess an identical delay of $\alpha$, then they will both move to $y(t=\alpha) = y_-$. However, at this point both agents once again occupy the same y-position, and will switch to $y(t=2\alpha) = y_+$, leading to an oscillation from which neither object will escape until collision.

It is also interesting to consider the cases when Alice and Bob have differing delays. In particular, let us analyze the scenario $\alpha_A < \alpha_B$. In this case, Alice will be the first to move to $y_-$ at $t = \alpha_A$. For any oscillation to occur, Bob's evaluation of the average relative distance must yield a value smaller than $R$. Thus, for oscillation to begin, we have the constraint

$$R > \frac{1}{\alpha_B}\int_{t-\alpha_B}^{t}|y_A[t] - y_B[t]| = 2h\left(\frac{\alpha_B - \alpha_A}{\alpha_B}\right) \tag{5}$$

Assuming this initial condition is met, and that Alice continues to evade with each subsequent step, then we can calculate the number of maneuvers executed by Alice and Bob. After Alice moves at $t = \alpha_A$, Bob will continue on his original track for another $\alpha_B - \alpha_A$, at which point the decision inequality for Bob is no longer satisfied. In this simplistic model, for each maneuver made by Bob, the systems are delayed by an additional time-difference of $\alpha_B - \alpha_A$. The system will cease to oscillate when $R > 2h\frac{n(\alpha_B - \alpha_A)}{\alpha_B}$, where $n$ is the number of maneuvers that have been executed by Alice. Thus, the system supports oscillatory behavior while the following inequality holds true:

$$n < \frac{R}{2h}\left(\frac{\alpha_B}{\alpha_B - \alpha_A}\right) \tag{6}$$

where Bob has made $n-1$ maneuvers.

The caveat to this simple result is the scenario where there exists an integer $n$ such that $\frac{R\alpha_A}{2h(\alpha_B - \alpha_A)} < n < \frac{R\alpha_B}{2h(\alpha_B - \alpha_A)}$. Under these conditions Bob switches too slowly for Alice's evasion condition to be met during that time step, despite not having reached avoidance, and it takes Alice until the next time step to maneuver away. This is an artifact of setting hard bounds on un-weighted decision time-integrals, which we have chosen for the sake of simplicity. On the other hand, the system is no longer in regime of the simple one-to-one stepping behavior we wished to examine in this paper, so to circumvent this we only chose parameters such that:

$$\frac{R\alpha_B}{2h(\alpha_B - \alpha_A)} - floor\left[\frac{R\alpha_A}{2h(\alpha_B - \alpha_A)}\right] > 1 \tag{7}$$

*2.3 Ternary model*

The ternary model allows for three possible courses which the agents may travel along: $y = 0, \pm h$. It represents a system where the agents have more room to perform an avoidance maneuver than in the binary model. Here, for simplicity, we will assume that the time delays of the agents are identical, and again, that the sidestepping action itself occurs instantaneously.

The two agents are on a collision course at $y(t=0) = 0$. However, now instead of deterministically evading to $y = h$, we allow a random decision between the two courses. If, after the first time-step, one agent evades to $y = -h$, while the other goes to $y = h$, they are no longer on a collision course, and thus their avoidance maneuvers are finished. Therefore, there is a 50% chance that no oscillation will occur in the system. Assuming that an oscillation has been initiated, both agents will deterministically return to $y = 0$, and once again make a random decision to evade in either direction. Within the parameters of this model, it is simple to calculate the probability distribution of the number of oscillations completed before avoidance is reached. With each completed oscillation, there is a 50% chance that the parties will successfully complete their maneuver. Thus, the probability of reaching $n$ oscillations within this model is an exponentially decaying function

$$P(n) = (0.5)^n \tag{8}$$

## 3. Computer simulations

The binary and ternary models were realized by computer simulations. Both models behave as expected under the imposed specifications, and follow quantitatively the models above.

*3.1 Binary model*

For the purposes of computer simulation, time is discretized, and the parameters of the binary model are set to $h = 1$ and $R = 0.5$. In order to set Alice and Bob on a collision course, we begin with both at $y = y_+$:

$$y_A[t = 0] = 1 , \quad y_B[t = 0] = 1 \tag{9}$$

In addition, since in our simulation the time and $x$-position values are restricted to discrete values, the time average in the inequality for our switching condition becomes

$$R > \frac{1}{\alpha_{AB}} \sum_{t-\alpha_{AB}}^{t} |y_A[t] - y_B[t]| \tag{10}$$

In the case of $\alpha_A = \alpha_B = 100$, the two agents are identical, and the binary system enters a perfect oscillation as shown in Fig 4: (a) from which the agents do not escape. We can see from the simulation that the agents simultaneously switch with a frequency of $\alpha$, as expected from our model. Even though time and space are discrete, the integral in equation (2) will give the same result as the sum in (5), because we have assumed that switching is instantaneous, yielding trajectories that are step-functions in $y$. Simulations of the binary model were also performed with an increasing delay difference between Alice and Bob. Fig. 4 (b) shows the case where a finite number of oscillations occur. We also know from the binary model that at a certain point Alice will be able to avoid Bob without Bob having to maneuver at all, which is demonstrated in Fig. 4: (d). We can find this limit by solving the inequality given by equation (3), which yields $\Delta\alpha_0 = \frac{R}{2h}\alpha_B$ where $\Delta\alpha_0$ is the minimum difference in delay times where Bob does not deviate from his original course. The total number of maneuvers performed is still given by equation (4). In Fig. 5 we have plotted the number of maneuvers performed by Bob with a constant time-delay of $\alpha_B = 100$ versus the time-delay $\alpha_A$. We can see that the functional relationship matches exactly our prediction from equation (4), and that the zero-oscillation point for $R = 0.5$ and $h = 1$ is $\Delta\alpha_0 = 25$, confirming our quantitative prediction.

(a)

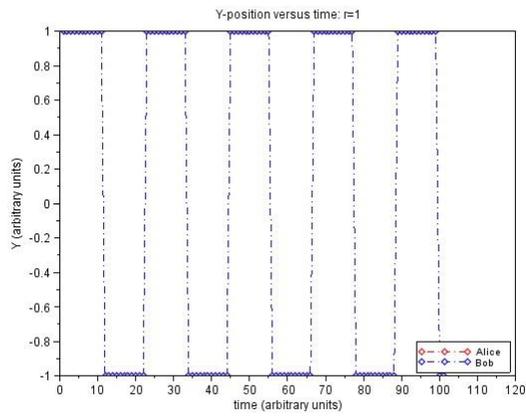

(b)

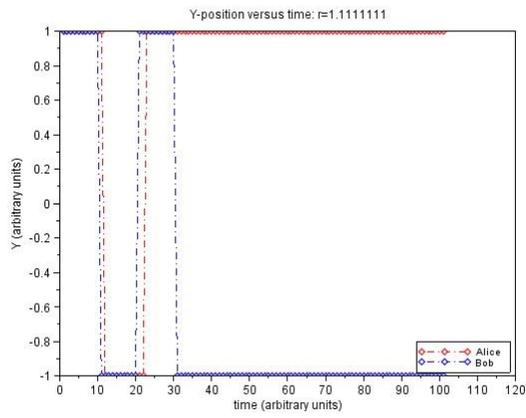

(c)

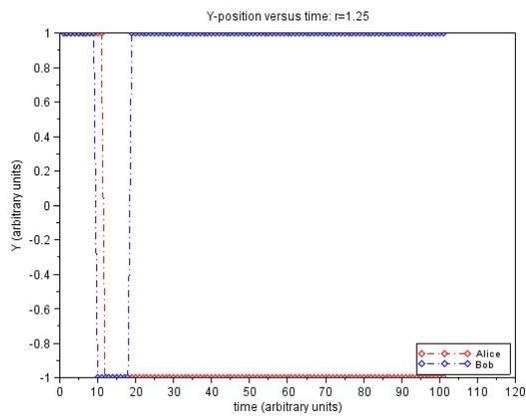

(d)

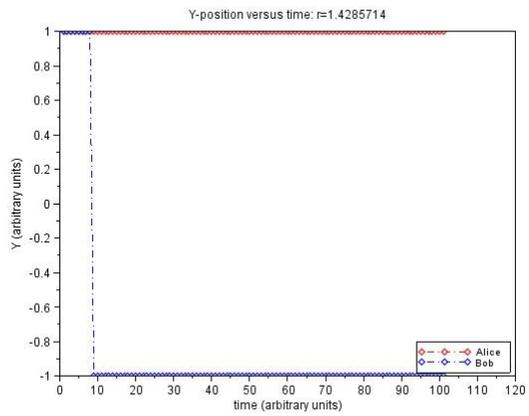

Figure 4. Binary model trajectories for increasing ratio $r = \frac{\alpha_A}{\alpha_B}$.

*3.2 Ternary model*

Within the ternary model, each return to the null position represents a nonzero probability of exiting the oscillation and completing the avoidance maneuver. Fig. 5 displays the transient behavior of a single system with $\alpha_A = \alpha_B = 10$.

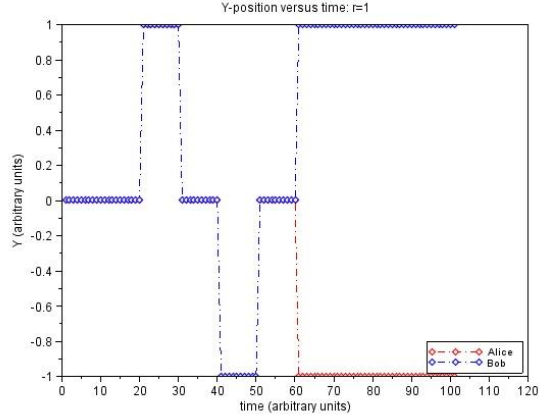

.
Figure 5. Ternary model trajectory with ratio $r = \frac{\alpha_1}{\alpha_2} = 1$.

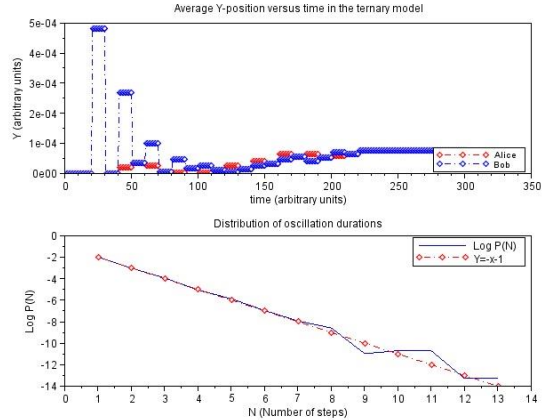

Figure 6. Ensemble-averaged ternary model trajectory and oscillation lifetime probability density.

As stated in equation (5), the probability of reaching $n$ oscillations decays as $P(n) = (0.5)^n$ .

By sampling the system numerous times, we attained a decaying exponential probability distribution, as can be seen in Fig. 6, which agrees with our calculation. Thus, we see that when the complexity of the system is increased by adding a statistical element, the additional room for deviation from oscillatory conditions can drastically reduce the lifetime of this phenomenon.

**4. Conclusion**

In conclusion, we have constructed the simplest models which will exhibit oscillatory behavior in an avoidance control problem. We have calculated the total number of evasive actions necessary in the binary problem, as well as the distribution for the lifetimes of oscillatory behavior in the ternary regime. Computer simulations were carried out to model these scenarios, and results agreed with the analytical calculations. Increasing the degrees of variability in the system produce a quicker degeneration of oscillatory behavior. Optimal strategies for the minimization of this effect could be the objects of future studies.

**References**

[1]  M.H. Breitner, ,"The Genesis of Differential Games in Light of Isaacs' Contributions", *Journal of Optimization Theory an Applications* **124** (2005) 523-559.

[2]  G. Leitmann and J. Skowronski, (1977). Avoidance Control. *Journal of Optimization Theory and Applications* **23** (1977), 581-591.

[3]  T. Tarnopolksaya and N. Fulton, (2009). Optimal cooperative Collision Avoidance Strategy for Coplanar Encounter: Merz's Solution Revisited. *Journal of Optimization Theory an Application* **140** (2009*)* 523-559.

[4]  N.E. Boudette, "Autopilot Cited in Death of Chinese Tesla Driver", *New York Times* (September 14, 2016). http://www.nytimes.com/2016/09/15/business/fatal-tesla-crash-in-china-involved-autopilot-government-tv-says.html

[5]  B. Vlasic and N.E. Boudette, "Self-Driving Tesla Was Involved in Fatal Crash, U.S. Says", *New York Times* (June 30, 2016). http://www.nytimes.com/2016/07/01/business/self-driving-tesla-fatal-crash-investigation.html

[6]  A. Narula-Tam, K. Nadumuri, S. Chaumette, D. Giustiniano, "Enabling next generation airborne communications", IEEE Communications Magazine **52** (2016), 102-103.

[7]  M. Schnell, U. Epple, D. Shutin, and N. Schneckenburger. "LDACS: Future aeronautical communications for air-traffic management." *IEEE Communications Magazine* **52** (2014) 104-110